# Full Electrical Switching of a Freestanding Ferrimagnetic Metal for Energy-Efficient Bipolar Neuromorphic Computing

*Li Liu[1,5], Peixin Qin[1,5],\*, Xiang Wang[2], Xiaobo She[2], Shaoxuan Zhang[3], Xiaoning Wang[1], Hongyu Chen[1], Guojian Zhao[1,5], Zhiyuan Duan[1,5], Ziang Meng[1,5], Qinghua Zhang[4],\*, Qiong Wu[3],\*, Yu Liu[2],\*, Zhiqi Liu[1,5],\**

[1]School of Materials Science and Engineering, Beihang University, Beijing 100191, China

[2]College of Integrated Circuit Science and Engineering, Nanjing University of Posts and Telecommunications, Nanjing 210023, China

[3]School of Materials Science and Engineering, Beijing University of Technology, Beijing 100124, China

[4]Beijing National Laboratory for Condensed Matter Physics, Institute of Physics, Chinese Academy of Sciences, Beijing 100190, China

[5]State Key Laboratory of Tropic Ocean Engineering Materials and Materials Evaluation, Hainan 570228, China

\*Corresponding authors. Emails: qinpeixin@buaa.edu.cn; zqh@iphy.ac.cn; wuqiong0506@bjut.edu.cn; y-liu-17@tsinghua.org.cn; zhiqi@buaa.edu.cn

**KEYWORDS**

**ABSTRACT**

Flexible electronics and neuromorphic computing face key challenges in material integration and function retention. In particular, freestanding membranes suffer from slow sacrificial layer removal and interfacial strain, while neuromorphic hardware often relies on area-intensive dual-device schemes for bipolar synaptic weights. Here, we present a universal strategy based on water-soluble $Sr_4Al_2O_7$ sacrificial layers, enabling the rapid release of freestanding ferrimagnetic metal membranes, which exhibit deterministic spin–orbit torque switching characteristics with well-preserved perpendicular magnetic anisotropy and are potential for next-generation ultrafast information technology. Extending this approach, we realize single-device ferrimagnetic synapses exhibiting intrinsic bipolar resistive switching. When implemented in a ResNet-18 architecture, these devices achieve 92% accuracy on CIFAR-10—comparable to floating-point software models—while halving device counts relative to differential-pair implementations. These results establish a scalable platform linking flexible spintronics with compact, high-performance neuromorphic systems, offering foundational advances for next-generation electronics and brain-inspired hardware.

**MAIN TEXT**

Freestanding membranes, defined as substrate-independent materials capable of maintaining structural integrity and intrinsic physical/chemical functionalities after detachment from their native growth substrates via specific techniques (*e.g.*, sacrificial layer exfoliation, interfacial assembly),[1] offer unique advantages by circumventing limitations imposed by traditional epitaxial constraints. These include enhanced strain engineering capabilities,[2-7] flexibility for multifunctional integration,[2,3,8-10] emergent quantum phenomena,[2,11] extreme environmental adaptability,[5,10] and unprecedented stacking/re-growth potential.[2,5,10,12,13] These characteristics establish freestanding membranes as transformative platforms for advanced functional material design and hybrid system integration.

Nevertheless, the exfoliation and transfer of freestanding membranes constitute a pivotal challenge in realizing functional heterostructures and flexible electronic devices. The isolation of freestanding membranes has followed a developmental trajectory rooted in two-dimensional material research. Since Geim and Novoselov first demonstrated monolayer graphene acquisition through Scotch tape-based mechanical exfoliation,[14] this methodology has been extended to other layered materials including $MoS_2$[15] and $WS_2$[16]. Despite inherent limitations in production yield and reproducibility, mechanical exfoliation remains the predominant laboratory technique for obtaining high-quality two-dimensional materials due to its operational simplicity and broad applicability. However, the strong interfacial coupling between conventional epitaxial thin films and rigid substrates fundamentally limits the effectiveness of mechanical exfoliation methods in preserving film integrity and crystallinity.[1,17,18] Recent developments in freestanding membrane fabrication have focused on alternative strategies including laser lift-off[19-21] and chemical etching.[17,22] Among these, the water-soluble sacrificial layer etching method has received significant attention.[1,3,12,23]

Notably, cubic $Sr_3Al_2O_6$ (SAO$_C$) sacrificial layers, leveraging their water solubility and perovskite-compatible pseudocubic lattice constants ($a^* \sim 3.96$ Å; corresponding cubic cell $a = 4 \times a^* \approx 15.84$ Å),[1,6,12,23-25] have enabled non-destructive exfoliation of single-crystal oxide films (*e.g.*, $BiMnO_3$,[25] $La_{0.7}Sr_{0.3}MnO_3$,[12,23] $SrTiO_3$,[23] and $BaTiO_3$[6,24]) without toxic etchants. However, the cubic phase structure of SAO$_C$ induces progressive interfacial stress accumulation during epitaxial growth, ultimately leading to crack formation that restricts attainable membrane dimensions.[24] Furthermore, the sluggish dissolution kinetics of SAO$_C$—requiring multiple hours for millimeter-scale film separation—renders it impractical for wafer-level manufacturing applications.[1,26]

It is worth noting that the most recent achievement of $Sr_4Al_2O_7$ (SAO$_T$) through epitaxial strain engineering marked a significant advance in oxide electronics.[1,3] In contrast to SAO$_C$, the low-symmetry crystalline architecture of SAO$_T$ (where biaxial strain converts orthorhombic to tetragonal symmetry) and its broad tunability in pseudocubic in-plane lattice constants of SAO$_T$ (experimental $a^* \sim 3.81$–$4.04$ Å; DFT-calculated $a^* = b^* = [(a^2 + b^2)^{1/2}]/4 \approx 3.896$ Å), alongside an out-of-plane $c^* = c/6 \approx 4.288$ Å, provide enhanced lattice adaptability to perovskite oxides, enabling broader compatibility with diverse oxide film systems.[1,27]

Crucially, SAO$_T$ achieves enhanced functionality through its discontinuous Al-O network and Sr-O-rich chemistry, accelerating water dissolution rates tenfold versus SAO$_C$[3], while its reduced symmetry concurrently suppresses interfacial strain. These combined attributes establish SAO$_T$-assisted transfer as the most promising approach for scalable production of freestanding membranes.

Despite these advances, water-soluble sacrificial layers (*e.g.*, SAO$_T$) remain predominantly applied to perovskite oxides (ABO$_3$-type), with limited exploration of metal film fabrication. This stems from tetragonal symmetry of SAO$_T$ aligning well with perovskites (*e.g.*, SrTiO$_3$, LaNiO$_3$) for coherent epitaxy[1], whereas metal thin films (*e.g.*, Cu, Al, Au) typically exhibit face-centered cubic or body-centered cubic structures, which differ significantly from the tetragonal symmetry and lattice constants of SAO$_T$. This structural mismatch makes it challenging to form a low-defect coherent interface, resulting in stress concentration and fracture susceptibility during the delamination process. Further challenges include oxidation risks during metal film processing.

Due to their antiferromagnetic exchange coupling, ferrimagnets exhibit ultrafast THz spin dynamics and can be electrically manipulated, making them promising for next-generation ultrafast information technologies.[28] As a typical ferrimagnet, CoTb alloys are widely studied in spintronics for their high spin polarization, strong perpendicular magnetic anisotropy (PMA), and tunable exchange coupling.[29,30] Concurrently, the intrinsic bipolar resistive behavior of CoTb ferrimagnetic materials enables controllable resistance state modulation in single-device configurations via spin-orbit torque (SOT) effects—a process that theoretically emulates the excitatory and inhibitory synaptic weight adjustments in biological systems. This characteristic positions CoTb as an ideal candidate for realizing single-device positive-negative weight representation, addressing the limitations of conventional memristors that typically exhibit unipolar conductance and necessitate differential pair designs (two devices per synapse), thereby enhancing the efficiency of neural network computations. However, realizing reliable devices faces challenges in interface control, strain management, and substrate compatibility. Fabricating freestanding CoTb membranes could overcome these limitations by decoupling material growth from device constraints, thereby facilitating optimized synthesis and functional development.

In this study, we integrate $Sr_4Al_2O_7$ sacrificial layers with amorphous ferrimagnetic CoTb films, thereby utilizing rapid water dissolution for $SAO_T$ to achieve Pt/CoTb freestanding films with over 200 μm dimensions. Systematic electrical transport characterization revealed the retention of PMA and demonstrated SOT-driven magnetization switching in processed devices. Furthermore, the unique potential of the underlying physical mechanisms of SOT-driven switching in neuromorphic computing was elucidated. Notably, the SOT-induced ferrimagnetic switching not only ensures rapid and energy-efficient synaptic updates but also eliminates the hardware redundancy inherent to conventional memristor differential pair designs through its intrinsic bipolar resistive behavior, thereby reducing synaptic unit counts by 50%. These advantages were validated by constructing a complete ResNet-18 neural network, where the ferrimagnetic devices were employed as synaptic weights for the CIFAR-10 image classification task. The hardware-emulated system achieved a classification accuracy of 92%, comparable to ideal floating-point implementations, highlighting the exceptional linearity and stability of the devices enabled by the advanced fabrication strategy. This direct mapping from spintronic physical mechanisms to neural network functional requirements provides dual innovations at both material and device levels, paving the way for energy-efficient and high-density neuromorphic hardware.

The $SAO_T$ epitaxial films were grown on MgO(001) substrates via pulsed laser deposition, serving as sacrificial templates for subsequent device fabrication (Figure 1a). Pt/CoTb bilayers were deposited on $SAO_T$ using magnetron sputtering under ultrahigh vacuum conditions (Figure 1b and 1c). Freestanding Pt/CoTb membranes were released by dissolving the $SAO_T$ layer and transferred onto Polydimethylsiloxane (PDMS) supports using a precision transfer system (Figure 1d-1f). Subsequently, Hall bar patterns with 5-μm-width channels were fabricated via UV lithography and Ar ion-beam etching (Figure 1g). Electrical contacts were formed by electron-beam evaporated Au/Cr electrodes (Figure 1h).

Figure 2a presents the XRD structural characterization of Pt/CoTb/SAO$_T$/MgO multilayer thin films. As shown in the pattern, the SAO$_T$ layer deposited at 700 °C demonstrates epitaxial growth on the (001)-oriented MgO substrate, evidenced by the SAO$_T$ (0012) diffraction peak at $2\theta \approx$ 42.68°. This peak position aligns with both the reported single-crystal SAO$_T$ (0012) orientation and the lattice expansion trend, where the SAO$_T$ lattice constant scales proportionally with the substrate[1,3,27]. These results conclusively demonstrate that our optimized deposition parameters (temperature and laser energy) enable phase-pure SAO$_T$ growth without secondary phases that could degrade chemical dissolution efficiency or promote microcrack propagation.[27] This phase stability is critical for ensuring structural integrity during film exfoliation, transfer procedures, and subsequent electrical characterization.

Figure 2b shows cross-sectional high-angle annular dark-field scanning transmission electron microscopy (HAADF-STEM) images of the as-grown Pt/CoTb/SAO$_T$/MgO heterostructure. The measured thicknesses of the Pt, CoTb, and SAO$_T$ layers are approximately 9 nm, 6 nm, and 175 nm, respectively, while the heterostructure exhibits remarkable interfacial planarity. Nevertheless, the rapid hydrolysis propensity of SAO$_T$ precludes direct observation of the (0012)-oriented epitaxial SAO$_T$ single-crystalline layer. The phase purity and structural integrity, however, were unambiguously validated by correlating the hydrolysis kinetics with single-crystal XRD characterization. Notably, the CoTb layers deposited at 100 °C exhibited no discernible crystalline lattice structure and were predominantly amorphous. The absence of long-range ordering likely circumvented grain boundary-induced crack propagation, which may account for the observed reduction in crack formation. The partial crystallization observed in the Pt layer could be attributed to its inherent high diffusion capability combined with bombardment effects from high-energy particles during sputtering.[31]

Presented in Figure 2c and 2d are high-resolution bright-field scanning transmission electron microscopy (BF-STEM) images of the freestanding Pt/CoTb bilayer, along with the corresponding fast Fourier transform (FFT) and inverse fast Fourier transform (IFFT) patterns. These images reveal a partially crystalline structure that aligns with the cross-sectional view in Figure 2b and is predominantly composed of Pt. The darker zones are attributed to regions of Pt aggregation, resulting from its tendency to agglomerate during sputtering. In contrast, the brighter zones also display crystallinity, indicating excellent continuity of the freestanding film; their enhanced brightness stems from reduced Pt coverage. As revealed by the energy-dispersive X-ray spectroscopy (EDS) spectra in Figure 2e, the freestanding Pt/CoTb bilayer demonstrates complete removal of the sacrificial layer $SAO_T$ along with high flatness in the transferred films. This exceptional flatness has provided optimal conditions for subsequent device fabrication and testing.

Figure 3a shows the Hall resistance characteristics of a Pt (9 nm)/CoTb (6 nm)/MgO heterostructure. The Hall loops exhibit well-defined squareness with an anomalous Hall resistance of ~ 0.075 Ω and pronounced PMA attributed to strong $3d$-$4f$ orbital hybridization.[29,32] Notably, the magnetization compensation temperature ($T_M$) resides in the 250-300 K range near room temperature, suggesting potentially enhanced SOT switching efficiency.[33,34]

Following successful transfer of the freestanding Pt/CoTb film onto a fresh MgO substrate via water-soluble $SAO_T$ sacrificial layer dissolution and PDMS-assisted transfer protocol, we implemented defect-mitigation strategies for electrical characterization. Photolithographic patterning coupled with anisotropic etching was employed to isolate a representative 370 × 200 μm² region (Figure 3b), creating precisely defined electrode fabrication zones. Sequential processing involving spin-coating, photolithography, and electron-beam evaporation enabled the construction of optimized Au/Cr bilayer electrodes. This optimized metallization process enhanced electrical interfacial properties and effectively preserved the integrity of freestanding Pt/CoTb stack by mitigating processing-induced damage. Post-transfer analysis (Figure 3c) confirms retention of PMA characteristics along with invariant Hall resistance and $T_M$ values, demonstrating structural and functional consistency throughout the SAO$_T$-assisted freestanding fabrication process. More importantly, this submillimeter-scale (100-300 μm), damage-free transfer methodology establishes a technological platform for implementing SOT-driven magnetization switching in freestanding magnetic heterostructures.

Through further processing of the freestanding Pt/CoTb thin films, we fabricated Hall bar devices with 5 μm-wide channels and deposited Au/Cr electrodes at the six channel terminals using the same methodology for SOT switching characterization (Figure 4a). The fundamental principle of SOT-driven magnetization reversal involves the conversion of charge current into spin current via the spin Hall effect in the heavy metal Pt layer, thereby inducing magnetic moment switching in the CoTb layer.[35-37] To ensure switching efficiency, we first aligned the magnetic moments of the ferrimagnetic layer by applying a 500 mT out-of-plane magnetic field prior to current injection. Pulse currents ranging from +20 mA to -20 mA (corresponding to current density $J_C \sim 4.4\times10^7$ A/cm²) were subsequently applied through the 5 μm-wide Hall bar channel. This performance is competitive with other high-performance spintronic materials, including Pt/Co[38] and W-V/CoFeB bilayers[39]. Concurrently, an in-plane magnetic field of 25 mT was applied along the current direction, with Hall resistance measurements conducted using a 100 μA probe current.

The SOT switching characteristics presented in Figure 4b and 4c demonstrate nearly 100% switching efficiency, as evidenced by the total resistance change consistent with that shown in Figure 3c. Deterministic magnetization reversal was achieved through polarity switching of the applied magnetic fields, exhibiting performance comparable to those observed in Pt/CoTb structures directly grown on MgO substrates, which is free from the constraints of the substrates and differs from the previous important work on SOT-driven switching in Pt/CoTb multilayers deposited on flexible polyimide substrates.[40] This successful demonstration of SOT switching in a freestanding metal film, a rarity due to the challenges of structural mismatch and oxidation, overcomes a critical barrier for the field. These results validate the potential of ferrimagnetic CoTb thin films for future flexible applications and integrated device implementations.

Figure 5 meticulously delineates the comprehensive process and outcomes of integrating our fabricated ferrimagnetic devices as synaptic weights within a ResNet-18 neural network for CIFAR-10 image recognition. Figure 5a illustrates the schematic of the ResNet-18 network architecture employed, which processes 32 × 32 pixel color images through multiple convolutional layers and a final fully connected layer to classify images into 10 categories. Crucially, all synaptic weights within the convolutional and fully connected layers of this network are directly replaced by arrays of our developed ferrimagnetic devices. Specifically, we extracted a series of resistance states from our device's Hall effect resistance loop under an in-plane magnetic field of -25 mT (purple circles in Figure 4c). Subsequently, a comparison between Figure 5b and 5c starkly highlights the significant advantages our novel devices offer at the hardware implementation level. Conventional memristor approaches, such as the typical one transistor-one resistor (1T1R) differential pair design depicted in Figure 5b, necessitate two devices to represent a single bipolar weight ($W = W^+ - W^-$) due to their typically unipolar conductance. This design paradigm not only substantially increases the complexity of peripheral control circuitry but also incurs additional hardware resource expenditure, where WL and BL denote word lines and bit lines, respectively, used for selecting specific memristor cells. In stark contrast, our innovatively designed devices (Figure 5c) ingeniously leverage their unique characteristic of exhibiting both increasing and decreasing resistance states under external stimuli, thereby enabling the direct and efficient realization of bipolar weights with a single device. This breakthrough design effectively halves the number of devices required per weight, leading to a substantial improvement in hardware resource utilization and integration density.

Following the validation of the device's feasibility as a synaptic weight, we conducted a systematic evaluation of the ResNet-18 network performance based on these devices. Figure 5d presents the evolution of training accuracy and loss curves as a function of training epochs. The plot clearly contrasts two scenarios: firstly, a software-based simulation employing ideal floating-point weights (accuracy and loss represented by red and blue solid lines, respectively), which denotes the theoretical optimal performance; and secondly, a hardware-emulated scenario using our experimentally characterized ferrimagnetic devices as weights (accuracy and loss shown as black and pink solid lines, respectively). Encouragingly, the accuracy and loss profiles achieved by the neural network incorporating our devices closely approximate those of the ideal floating-point implementation, exhibiting minimal performance degradation. Our ferrimagnetic synaptic devices deliver a competitive 92% accuracy on CIFAR-10, rivaling the performance of state-of-the-art neuromorphic systems. For instance, while binary networks such as IR-Net report 86.5% accuracy,[41] and recent in-memory computing systems including VC-SOT-MRAM CIM and flash-based blocks reach 91.46%[42] and 91.80%[43] respectively, our method attains a comparable level without relying on differential pairs, with a concomitant 50% reduction in synaptic components, and entirely avoids analog-to-digital conversion overhead. This level of fidelity is notably challenging to achieve with conventional synaptic devices, which often suffer from performance penalties. We attribute this remarkable performance primarily to the excellent linearity and material stability of our devices. This robustly demonstrates the capability of our devices to effectively represent and modulate synaptic weights, thereby supporting complex learning tasks.

To provide a more granular analysis of the network's classification efficacy across different categories, Figure 5e and 5f presents the confusion matrices for the CIFAR-10 test set, corresponding to the ideal floating-point weight network and our device-based weight network, respectively. A meticulous comparison of these matrices reveals a high degree of concordance in classification accuracy across all image categories between the two implementations. Further supporting this consistent performance throughout training, detailed comparative confusion matrices at intermediate stages, specifically Epoch 25 and Epoch 75, are provided in Figure S1 and Figure S2, respectively. For instance, for the 'airplane' category, the ideal model correctly classified 949 instances, while the device-based model correctly classified 920. Conversely, for other categories such as 'cat', the ideal model achieved 832 correct classifications, whereas our device-based model achieved 843, even outperforming the ideal model in certain specific categories. Despite these minor category-specific variations, the overall performance disparity remains negligible, further substantiating the robustness and efficacy of our developed devices in practical, complex application scenarios.

Finally, Figure 5g vividly illustrates the dynamic evolution of quantized weight distributions at various training stages (Epochs 0, 25, 50, 75, and 100) using histograms. This plot intuitively reveals the adaptive adjustment mechanism of weights during the learning process: in the initial training phase (Epoch 0), the weight distribution is relatively diffuse, reflecting the random initialization state. As training progresses, the network learns patterns from the data via backpropagation, causing the magnitudes of weights corresponding to important features to increase, while weights for irrelevant or redundant features tend towards zero or smaller values. This leads to a gradual concentration of the weight distribution around specific values, particularly near zero. This evolutionary trend of the weight distribution, from dispersed to concentrated, aligns closely with the intrinsic mechanisms of knowledge distillation and parameter optimization in neural network training, and underscores the capability of our devices to support such fine-grained weight adjustments. A detailed view of the final learned weight distributions for the initial convolutional layers at Epoch 100, exemplifying this characteristic sparse and bipolar nature, is presented in Figure S3.

In summary, this work demonstrates the integration of $SAO_T$ sacrificial layers with CoTb films to fabricate freestanding ferrimagnetic membranes for area-efficient bipolar synaptic weights in neuromorphic computing. The rapid dissolution and low interfacial strain of $SAO_T$ enabled transfer of freestanding Pt/CoTb membranes with suppressed grain-boundary defects, while retaining PMA and considerable SOT-driven switching efficiency.[32] Furthermore, the refined sacrificial layer-assisted transfer technique developed in this work has been extended to neuromorphic computing applications, enabling high-quality ferrimagnetic synapses with intrinsic bipolar resistive switching in a single-device configuration. This design halves the number of devices per synapse compared to conventional differential pairs, thereby doubling integration density and significantly improving energy efficiency, without compromising computational performance. The practical viability of these synaptic elements was demonstrated through integration into a ResNet-18 neural network for CIFAR-10 image classification, achieving an accuracy of 92% which is virtually on par with ideal software-based floating-point implementations. This material-to-system innovation—from the SAO$_T$-enabled freestanding membrane synthesis to the compact bipolar synaptic device and eventually to high-accuracy neural network emulation—establishes a scalable pathway toward energy- and area-efficient neuromorphic computing. Looking forward, this platform can be further extended to other freestanding material systems such as emerging altermagnets (*e.g.*, RuO$_2$)[44,45] and noncollinear antiferromagnets (*e.g.*, Mn$_3$Sn and Mn$_3$Pt)[46-48], offering enhanced switching dynamics and broader application prospects in real-time recognition systems and low-power biomedical implants.

## ASSOCIATED CONTENT

**Supporting Information**

The Supporting Information is available free of charge.

## AUTHOR INFORMATION

**Corresponding Author**

*Emails: qinpeixin@buaa.edu.cn; zqh@iphy.ac.cn; wuqiong0506@bjut.edu.cn; y-liu-17@tsinghua.org.cn; zhiqi@buaa.edu.cn

**Author Contributions**

The manuscript was written through contributions of all authors. All authors have given approval to the final version of the manuscript.

**Notes**

The authors declare no competing financial interest.

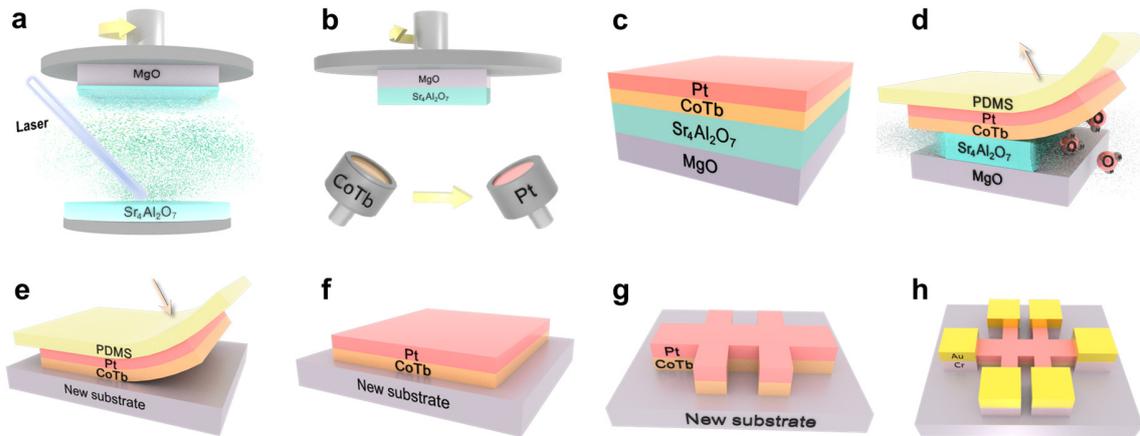

**Figure 1. Schematic of the process to obtain freestanding Pt/CoTb membranes and SOT-driven device fabrication.** (**a**) Deposition of the $Sr_4Al_2O_7$ ($SAO_T$) sacrificial layer via pulsed laser deposition. (**b**) Growth of the CoTb magnetic layer and the Pt heavy-metal capping layer via magnetron sputtering; (**c**) Structural overview of the Pt/CoTb/$SAO_T$ heterostructure. (**d**) Polydimethylsiloxane (PDMS)-assisted water dissolution of $SAO_T$ and exfoliation of freestanding Pt/CoTb membranes. (**e**) Transfer of freestanding Pt/CoTb membranes onto a new substrate. (**f**) Peeling off the PDMS support. (**g**) Patterning of Hall bar devices via UV photolithography and $Ar^+$ ion beam etching. (**h**) Deposition of Au/Cr bilayer electrodes by electron-beam evaporation.

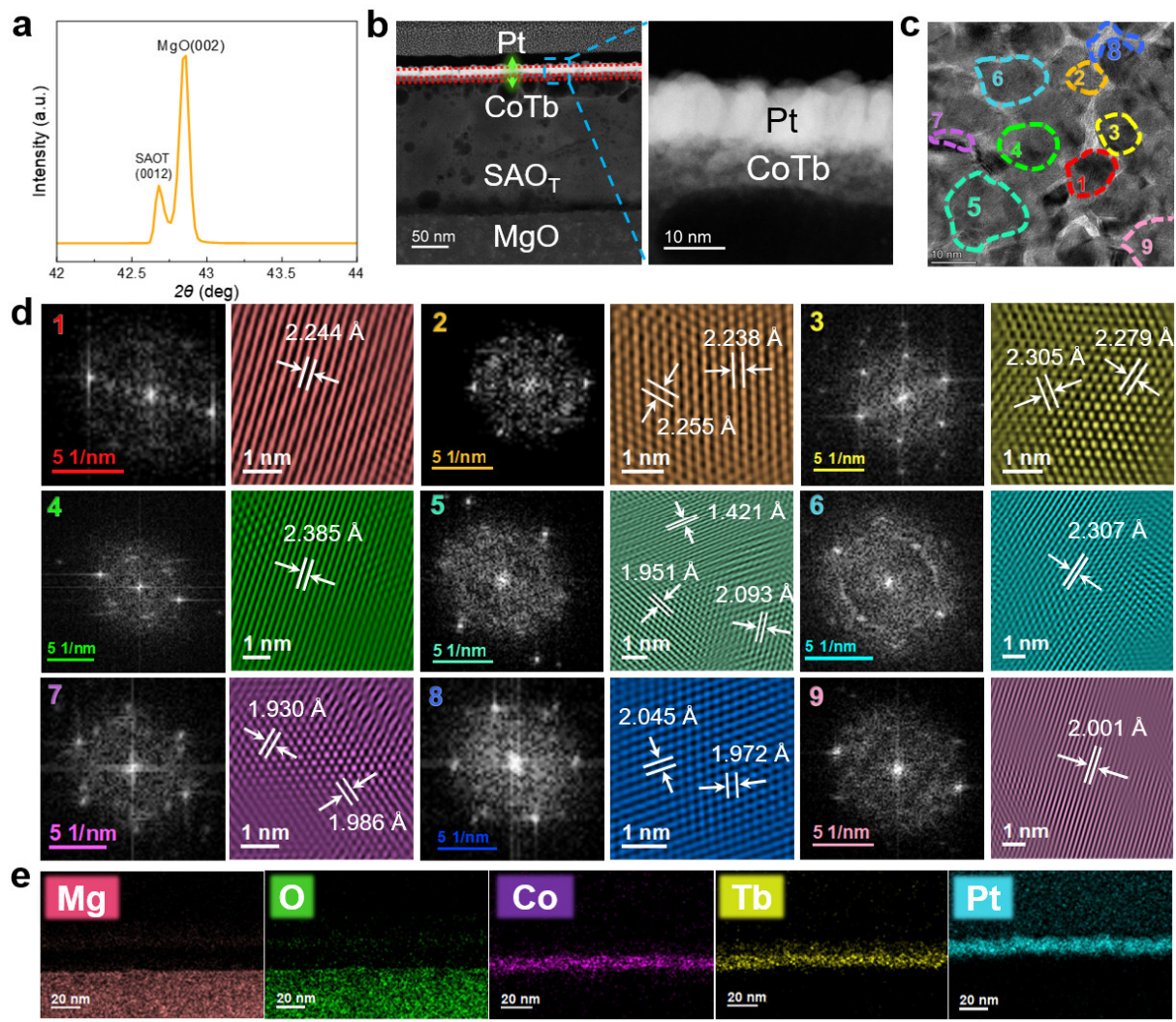

**Figure 2. Structural characterization.** (**a**) X-ray diffraction (XRD) patterns of the as-grown Pt/CoTb/Sr$_4$Al$_2$O$_7$(SAO$_T$) heterostructure, showing epitaxial (001)-oriented SAO$_T$ on MgO substrate. (**b**) Cross-sectional high-angle annular dark-field scanning transmission electron microscopy (HAADF-STEM) image of Pt/CoTb/SAO$_T$ and corresponding zoom in image of labeled regions. (**c**) High-resolution bright-field scanning transmission electron microscopy image of the freestanding Pt/CoTb bilayer. (**d**) The corresponding fast Fourier transform (FFT) and inverse fast Fourier transform (IFFT) images from the nine regions labeled in (c). (**e**) Energy-dispersive X-ray spectroscopy (EDS) spectra of the cross-sectional freestanding Pt/CoTb bilayer.

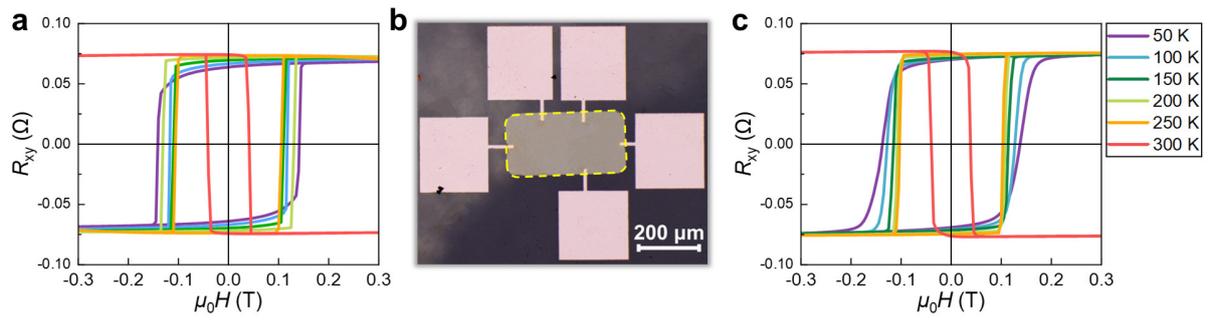

**Figure 3. Hall effect and morphology of Pt/CoTb films.** (**a**) Hall effect of as-grown Pt/CoTb on MgO substrate. (**b**) Optical micrograph of the freestanding Pt/CoTb membrane. (**c**) Hall effect of the freestanding Pt/CoTb membrane.

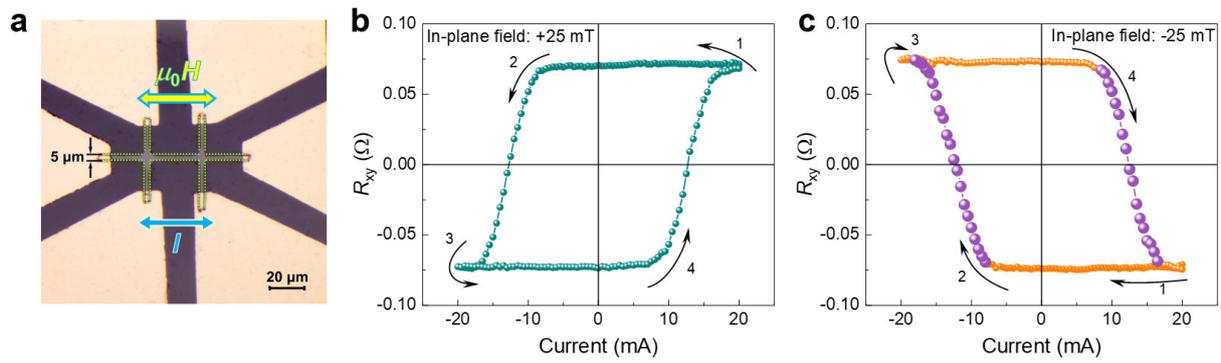

**Figure 4. Spin-orbit torque (SOT) switching of a freestanding Pt/CoTb membrane.** (**a**) Optical micrograph of the transferred freestanding Pt/CoTb Hall bar device; (**b, c**) Hall resistance versus current pulses under ±25 mT in-plane fields at 300 K. The purple balls represent the data points used for the neuromorphic computing.

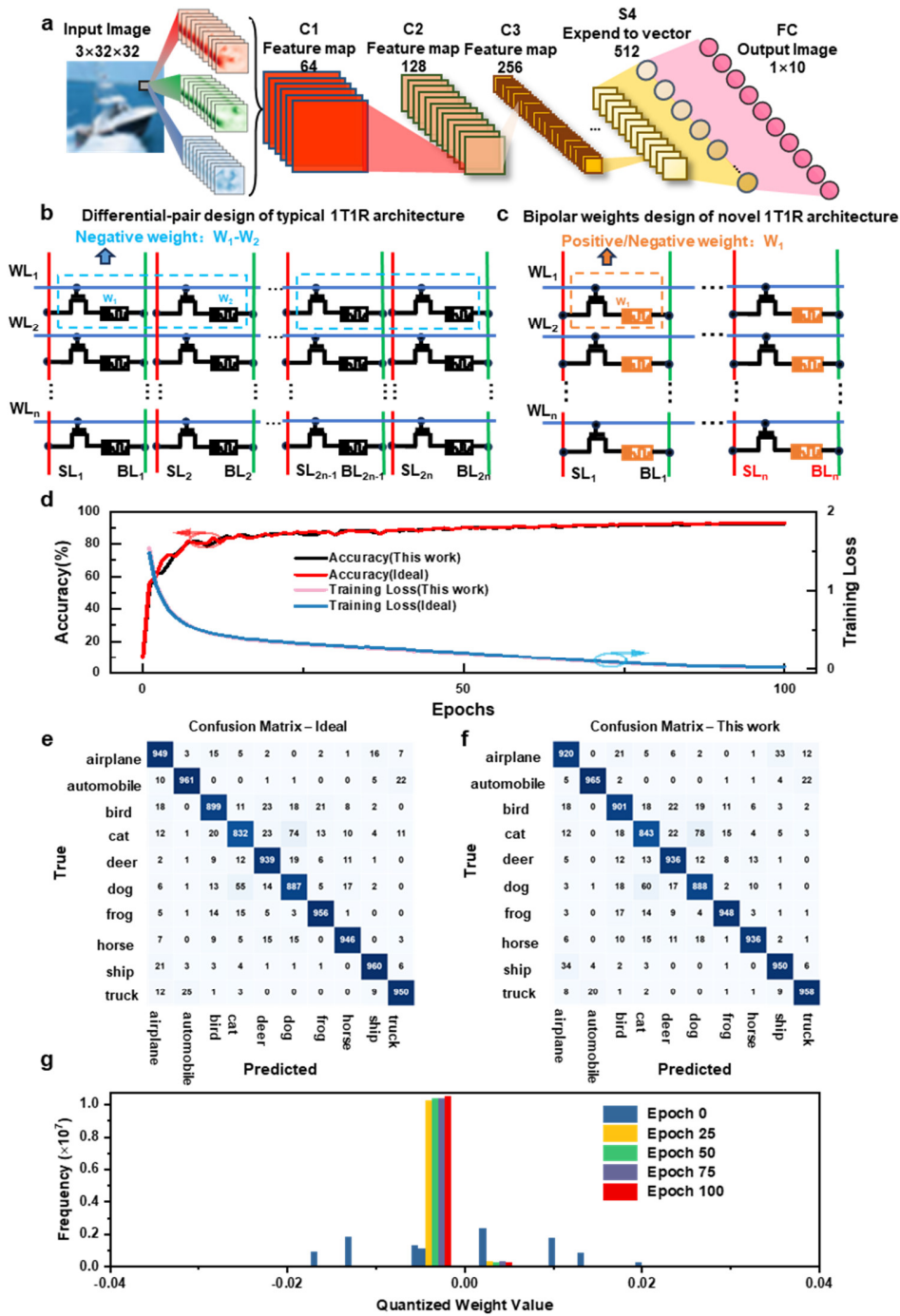

**Figure 5. ResNet-18 network implementation with ferrimagnetic synaptic devices and performance for CIFAR-10 image recognition.** (**a**) Schematic of the ResNet-18 architecture employed for CIFAR-10 image classification. The input is a 32 × 32 × 3 image, processed through convolutional and fully connected layers, with all synaptic weights implemented by the fabricated ferrimagnetic devices. (**b**) Illustration of a conventional differential-pair design using typical 1T1R unipolar memristors to represent a single bipolar synaptic weight, requiring two devices per

synapse. WL and BL denote word lines and bit lines, respectively. (**c**) Schematic of the proposed synaptic weight implementation using a single ferrimagnetic device capable of intrinsic bipolar weighting, thereby reducing hardware resource requirements by approximately 50% compared to the differential-pair approach. (**d**) Training accuracy and loss curves for the ResNet-18 network on CIFAR-10. Performance of the network with ferrimagnetic device weights is compared against an ideal software implementation with floating-point weights, demonstrating minimal performance degradation. (**e**) Confusion matrix for the CIFAR-10 test set using the ideal floating-point weight ResNet-18 network after 100 epochs. (**f**) Confusion matrix for the CIFAR-10 test set using the ResNet-18 network with ferrimagnetic device weights after 100 epochs, showing classification performance comparable to the ideal case. (**g**) Evolution of quantized weight distributions in the ferrimagnetic device-based network at different training epochs (0, 25, 50, 75, and 100), illustrating the dynamic adaptation of synaptic weights during the learning process.

**Full Electrical Switching of a Freestanding Ferrimagnetic Metal for Energy-Efficient Bipolar Neuromorphic Computing**

*Li Liu[1,5], Peixin Qin[1,5,6,*], Xiang Wang[2], Xiaobo She[2], Shaoxuan Zhang[3], Xiaoning Wang[1], Hongyu Chen[1], Guojian Zhao[1,5], Zhiyuan Duan[1,5], Ziang Meng[1,5], Qinghua Zhang[4,*], Qiong Wu[3,*], Yu Liu[2,*], Zhiqi Liu[1,5,*]*

[1]School of Materials Science and Engineering, Beihang University, Beijing 100191, China

[2]College of Integrated Circuit Science and Engineering, Nanjing University of Posts and Telecommunications, Nanjing 210023, China

[3]School of Materials Science and Engineering, Beijing University of Technology, Beijing 100124, China

[4]Beijing National Laboratory for Condensed Matter Physics, Institute of Physics, Chinese Academy of Sciences, Beijing 100190, China

[5]State Key Laboratory of Tropic Ocean Engineering Materials and Materials Evaluation, Hainan 570228, China

[6]Wuhan National High Magnetic Field Center, Huazhong University of Science & Technology, Wuhan 430074, China

*Experimental Methods*

*Film Growth*

Sr$_4$Al$_2$O$_7$ (SAO$_T$) epitaxial films were grown on (001)-oriented MgO substrates via pulsed laser deposition (KrF excimer laser) at 700 °C for 15 min with a per-pulse laser fluence of 1.6 J/cm$^2$. Pt/CoTb multilayers were then deposited by magnetron sputtering in an ultrahigh vacuum chamber (base pressure ≤ 7.5×10$^{-8}$ Torr). The CoTb alloy layer was co-sputtered under 3×10$^{-3}$ Torr argon atmosphere with DC power settings of 60 W (Co) and 30 W (Tb) for 2 min, immediately followed by a Pt capping layer deposited at 30 W DC for 4 min. Continuous substrate rotation ensured thickness consistency.

*Membrane Transfer*

Freestanding Pt/CoTb membranes were fabricated by dissolving the SAO$_T$ sacrificial layer in deionized water for 10 min, with the Polydimethylsiloxane (PDMS) support maintaining mechanical integrity during release. The membrane was transferred to a new MgO substrate using a precision alignment Metatest E1-G transfer system.

*Device Fabrication*

Hall bar devices with 5-μm channel widths were patterned through UV photolithography using AZ5214 photoresist spin-coated at 4000 rpm. The patterns were processed via Ar ion-beam etching (80 mA beam current, 250 s duration). Au/Cr bilayer electrodes (60/15 nm) were deposited by electron-beam evaporation, followed by lift-off in acetone using an ultrasonic bath.

*Characterization Techniques*

Electrical transport measurements were conducted in a Quantum Design VersaLab integrated with precision instrumentation: a Keithley 6221 current source provided excitation currents, while the Hall voltage was resolved using a Keithley 2182A nanovoltmeter. An external magnetic field was applied perpendicular to the film plane during anomalous Hall effect measurements. Cross-sectional microstructure evaluation was performed using a JEOL NeoArm transmission electron microscope operated at 200 kV. The XRD data was obtained with a Rigaku SmartLab diffractometer using Cu-$K\alpha_1$ radiation ($\lambda$ = 0.154056 nm). Surface morphology and device metrology were characterized through optical profilometry using a Metatest E1-G metallurgical microscope. The film stacks in this work are Pt (9 nm)/CoTb (6 nm) unless otherwise noted.

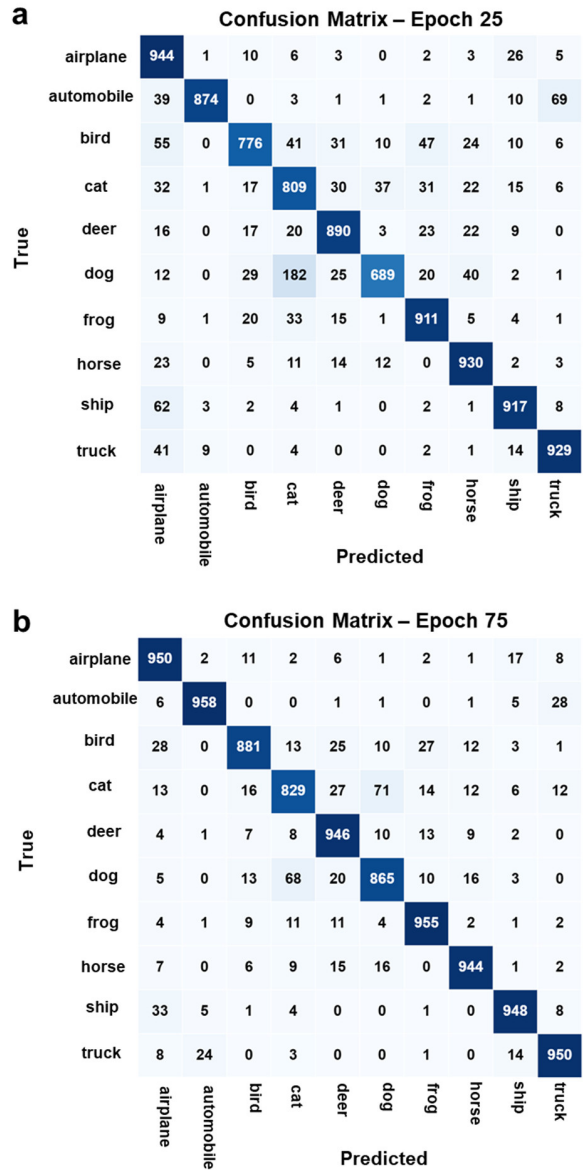

**Figure S1. Comparative confusion matrices for CIFAR-10 classification at training epoch 25.** Confusion matrices detailing the classification performance of the ResNet-18 network on the CIFAR-10 dataset after 25 training epochs. **(a)** Performance of the network utilizing ideal floating-point weights (software simulation). **(b)** Performance of the network implemented with weights mapped from the experimentally characterized ferrimagnetic devices. Each matrix plots true labels against predicted labels, providing a class-wise performance comparison at an early-to-intermediate training stage.

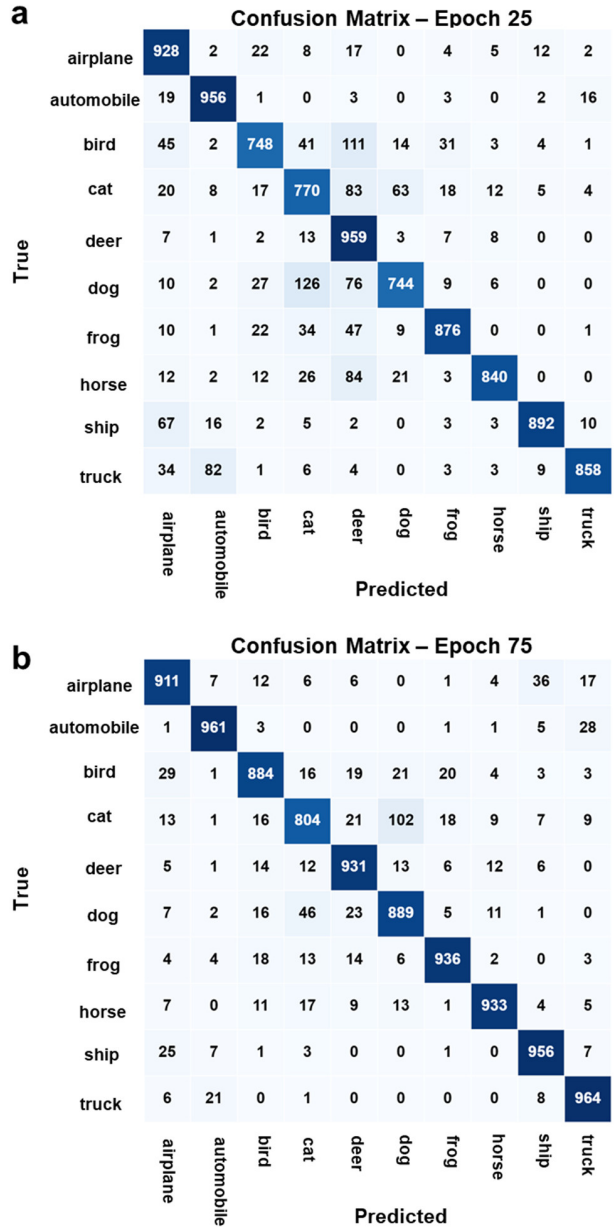

**Figure S2. Comparative confusion matrices for CIFAR-10 classification at training epoch 75.** Confusion matrices for the CIFAR-10 classification task after 75 training epochs. **(a)** Results for the ResNet-18 network with ideal floating-point weights. **(b)** Results for the ResNet-18 network employing the ferrimagnetic (FIM) device-based weights. This comparison at a later training stage further illustrates the consistent learning capability and sustained classification accuracy of the FIM device-based neural network relative to the ideal software model.

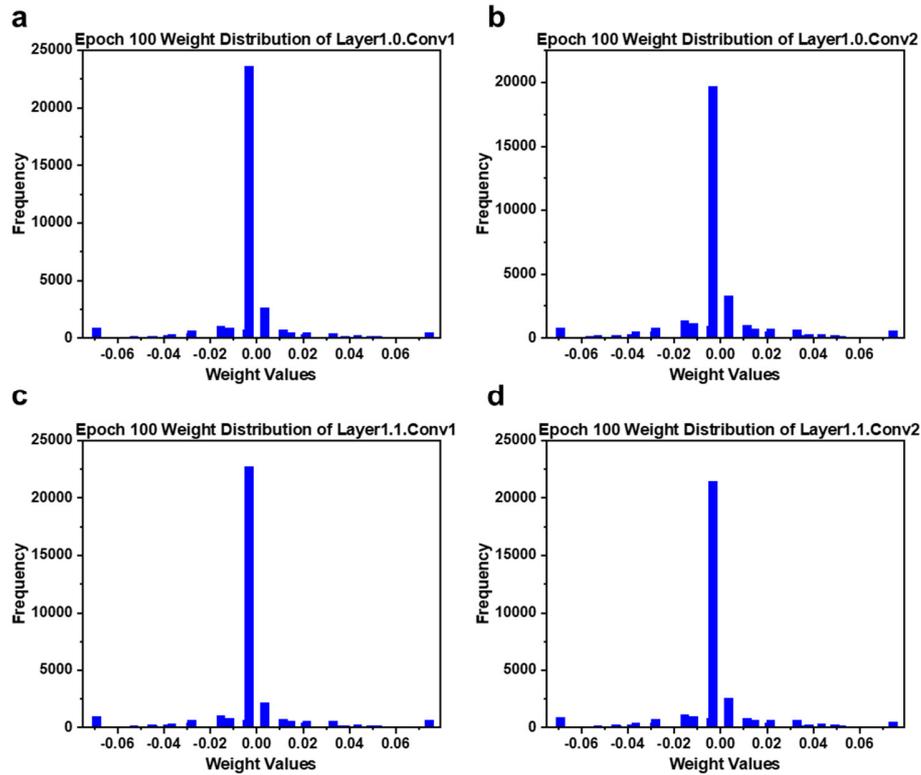

**Figure S3. Weight distributions in initial convolutional layers of the ferrimagnetic device-based network at epoch 100.** Histograms showing the distribution of quantized weight values for the first four convolutional layers of the ResNet-18 network after 100 training epochs. All weights are implemented using the characterized ferrimagnetic synaptic devices. The distributions demonstrate a characteristic learned state with a significant proportion of weights near zero and distinct populations of positive and negative weights, reflecting effective synaptic potentiation and depression.